%% file: paper.tex
\documentclass[ALICE,manyauthors]{cernphprep}

\usepackage[comma,square,numbers,sort&compress]{natbib}
\usepackage{hyperref}
\usepackage{lineno}
\usepackage{subfigure}
\usepackage{tikz}
\usepackage[section]{placeins}
\usepackage {multicol}
\usepackage {multirow}
\usepackage {units}
\usepackage {array}
\usepackage{verbatim}
\usepackage{dblfloatfix} 
\usepackage[T1]{fontenc}
\usepackage{orcidlink}
\usepackage[T1]{fontenc}
\usepackage{orcidlink}

\input{commands.tex}

\begin{document}

\begin{titlepage}
\PHyear{2024}       
\PHnumber{229}      
\PHdate{06 September}  

\title{Multiplicity-dependent jet modification from di-hadron correlations in \pp{} collisions at ${\sqrt{{\textit s}}}=\mathbf{13}$~TeV}
\ShortTitle{Multiplicity-dependent di-hadron correlations in \pp{} collisions}   

\Collaboration{ALICE Collaboration\thanks{See Appendix~\ref{app:collab} for the list of collaboration members}}
\ShortAuthor{ALICE Collaboration} 

\begin{abstract}
Short-range correlations between charged particles are studied via two-particle angular correlations in \pp{} collisions at ${\sqrt{{\textit s}}}=13$~TeV. The correlation functions are measured as a function of the relative azimuthal angle $\Delta\varphi$ and the pseudorapidity separation $\Delta\eta$ for pairs of primary charged particles within the pseudorapidity interval $|\eta| < 0.9$ and the transverse-momentum range $1 < \pt < 8$ GeV/$c$. Near-side ($|\Delta\varphi|<1.3$) peak widths are extracted from a generalised Gaussian fitted over the correlations in full pseudorapidity separation ($|\Delta\eta|<1.8$), while the per-trigger associated near-side yields are extracted for the short-range correlations ($|\Delta\eta|<1.3$). Both are evaluated as a function of charged-particle multiplicity obtained by two different event activity estimators. The width of the near-side peak decreases with increasing multiplicity, and this trend is reproduced qualitatively by the Monte Carlo event generators PYTHIA 8, AMPT, and EPOS.  However, the models overestimate the width in the low transverse-momentum region ($\pt < 3$ GeV/$c$). 
The per-trigger associated near-side yield increases with increasing multiplicity. Although this trend is also captured qualitatively by the considered event generators, the yield is mostly
overestimated by the models in the considered kinematic range.
The measurement of the shape and yield of the short-range correlation peak can help us understand the interplay between jet fragmentation and event activity, quantify the narrowing trend of the near-side peak as a function of transverse momentum and multiplicity selections in \pp{} collisions, and search for final-state jet modification in small collision systems.

\end{abstract}

\end{titlepage}

\setcounter{page}{2}

\input{sec_intro.tex}
\input{sec_ana.tex}

\input{sec_results.tex}
\input{sec_summary.tex}

\newenvironment{acknowledgement}{\relax}{\relax}
\begin{acknowledgement}
\section*{Acknowledgements}
\input{fa_2024-08-25_Opt_C.tex}    
\end{acknowledgement}

\bibliographystyle{utphys}
\bibliography{paper.bib}

\newpage
\appendix

\section{The ALICE Collaboration}
\label{app:collab}
\input{Alice_Authorlist_2024-08-25_Opt_C.tex}  

\end{document}

%% file: commands.tex
\newcommand{\AAA}          {AA}
\newcommand{\pp}           {pp}
\newcommand{\pA}           {pA}

\newcommand{\PbPb}         {\mbox{Pb--Pb}}
\newcommand{\pPb}          {\mbox{p--Pb}}
\newcommand{\AuAu}         {\mbox{Au--Au}}
\newcommand{\XeXe}         {\mbox{Xe--Xe}}
\newcommand{\CuCu}         {\mbox{Cu--Cu}}
\newcommand{\UrUr}         {\mbox{U--U}}
\newcommand{\ZrZr}         {\mbox{Zr--Zr}}
\newcommand{\RuRu}         {\mbox{Ru--Ru}}

\newcommand{\pt}           {\ensuremath{p_\mathrm{T}}}

\newcommand{\pTt}           {\ensuremath{p_{\rm{T,~trig}}}}
\newcommand{\pTa}           {\ensuremath{p_{\rm{T,~assoc}}}}

\newcommand{\snn}          {\ensuremath{\sqrt{s_{\rm NN}}}}

\newcommand{\pttrig}{\ensuremath{p_\mathrm{T,\,trig}}}
\newcommand{\ptassoc}{\ensuremath{p_\mathrm{T,\,assoc}}}



%% file: sec_intro.tex

\section{Introduction}
\label{sec:intro}

In high-energy particle collisions, measurements of particle correlations provide information on a wide range of physics effects, leading to quantitative and qualitative understanding of diverse phenomena in elementary (e.g.~electron--positron), small (proton--proton and proton--nucleus), and large (nucleus--nucleus) collision systems. In fact, one of the first surprising results from the Large Hadron Collider (LHC) physics programme was the conclusive observation of correlations in the transverse plane between particles across a wide range of pseudorapidity in \pp{} collisions~\cite{CMS:2010ifv}. This was reminiscent of similar long-range correlations that had previously been observed in nucleus--nucleus (\AAA{}) collisions, and which had been attributed to the expansion of a hot and dense strongly-interacting medium produced in such interactions, the quark--gluon plasma (QGP)~\cite{Adcox:2004mh,Adams:2005dq,Arsene:2004fa,Back:2004je,ALICE:2022wpn}. In heavy-ion collisions, measurements of correlations in momentum space give insight into both the collective \textit{bulk} dynamics of the system expansion as well as the interactions of hard and soft probes with the surrounding medium~\cite{Apolinario:2022vzg}. Searches for similar signals in small collision systems are an active area of research. 
Current research in small collision systems focuses on varying system sizes to explore and identify the onset of signatures that are attributed to the formation of a QGP~\cite{Nagle:2018nvi,Brewer:2021kiv}.

In heavy-ion collisions, two- and multi-particle angular correlation~\cite{Wang:1991qh,E877:1996czs,Borghini:2001vi,STAR:2002hbo} measurements demonstrate that the produced medium exhibits strong \textit{collectivity}~\cite{Ollitrault:1992bk}, in which many particles show correlated behaviour despite being far apart in (pseudo)rapidity. The anisotropies in the azimuthal-angle distribution of the emitted final-state particles have been extensively studied in nucleus--nucleus collisions from below 100$A$ MeV up to Super Proton Synchrotron (SPS) energies (see Refs.~\cite{Reisdorf:1997fx,Herrmann:1999wu,Voloshin:2008dg} and references therein); in \AuAu{}, \CuCu{}, \ZrZr{}, \RuRu{}, and \UrUr{} collisions in the centre-of-mass energy range $4.5 < \snn < 200$~GeV at the Relativistic Heavy Ion Collider (RHIC)~\cite{Adams:2005dq,Adcox:2004mh,Arsene:2004fa,Back:2004je,STAR:2020dav,PHENIX:2008osq,PHENIX:2006dpn,STAR:2015mki,STAR:2021mii}; and in \PbPb{} and \XeXe{} collisions at $\snn = 2.76$, $5.02$, and $5.44$~TeV at the LHC~\cite{Abelev:2012di, Abelev:2014pua, ATLAS:2011ah,ALICE:2022wpn,ALICE:2018lao}. The anisotropy in the final-state momentum-space particle distribution is commonly explained as emerging from the initial-state spatial anisotropy through the pressure-driven expansion of strongly interacting matter, which in the SPS/RHIC/LHC energy regime is expected to be the QGP. Models that describe the QGP expansion with hydrodynamic equations, such as iEBE-VISHNU~\cite{Shen:2014vra} and MUSIC~\cite{Schenke:2010nt}, are particularly successful in reproducing experimental measurements~\cite{Bernhard:2015hxa,Bernhard:2019bmu,Schenke:2010rr,Schenke:2020mbo} of \textit{anisotropic flow}. 
While hydrodynamic models suggest that anisotropic flow arises predominantly from final-state interactions, models of the initial conditions indicate that momentum-space anisotropies may also arise in the initial state. In particular, the interaction of dense colour fields in the colliding nuclei can produce a state known as the colour glass condensate (CGC) in the early moments of the collision~\cite{McLerran:1993ka,McLerran:1993ni,Iancu:2003xm}.  The resulting colour flux tubes may evolve to preferentially emit particles anisotropically in the azimuthal direction~\cite{Dumitru:2010iy,Dusling:2012iga}. To what extent these initial-state effects persist throughout the evolution of the collision, and contribute to the experimentally-observed final-state anisotropies, remains an open question.

Long-range correlations have been also observed in smaller collision systems such as proton--proton (pp)~\cite{CMS:2010ifv, ALICE:2023ulm,ATLAS:2015hzw, Khachatryan:2015lva, Khachatryan:2016txc, Acharya:2019vdf, ATLAS:2017rtr}, proton--nucleus (pA)~\cite{ALICE:2012eyl, ATLAS:2014qaj, ATLAS:2016yzd, Khachatryan:2016ibd}, and light nucleus--nucleus~\cite{PHENIX:2018lia, Aidala:2017ajz} collisions.  These results were surprising, as it was not expected that \pp{} and \pA{} collisions could produce the high energy density spread over a large spatial volume that is necessary for a QGP medium to form.
However, the origin of these long-range correlations in small systems is still not understood and is not clear whether they emerge due to the same underlying mechanisms as in the large collision systems.

Hydrodynamic models which were developed to describe heavy-ion collisions, such as superSONIC~\cite{Romatschke:2015gxa}, iEBE-VISHNU~\cite{Zhao:2020pty}, and MUSIC~\cite{Schenke:2020mbo}, also find success in describing the observed anisotropies in small collision systems.  Phenomenological Monte Carlo (MC) models based on Lund strings, such as PYTHIA~\cite{Bierlich:2022pfr, Skands:2014pea}, are able to generate \pp{} collisions with a large number of particles in the final state (high multiplicity) with colour reconnection, strangeness enhancement~\cite{ALICE:2016fzo} and isotropic (radial) expansion with colour rope fragmentation~\cite{Fischer:2016zzs, Ortiz:2016kpz}, and anisotropy in the final-state momentum distributions by interactions of colour strings via a mechanism called \textit{string shoving}~\cite{Bierlich:2017vhg}. However, they are not yet able to qualitatively describe the full range of QGP-like behaviour observed in small collision systems (see, for example, Ref.~\cite{ALICE:2021nir}). 
While the contributions from initial state momentum correlations described by gluon saturation models are expected to be more important in small collision systems than in large systems, initial state effects alone are not able to fully describe the systematic trends observed in experimental data (see, for example, Ref.~\cite{PHENIX:2018lia}). However, hybrid models comprising CGC-based initial state momentum correlations followed by a hydrodynamic evolution do provide good description of the data~\cite{Schenke:2020mbo}, and new differential observables are promising for disentangling initial- and final-state effects~\cite{Giacalone:2020byk}.  

Flow measurements in small collision systems present a significant experimental challenge due to the contribution of non-flow correlations (processes that produce correlated particles which are not collective, namely jet fragmentation, as described below). In contrast to the case of heavy-ion collisions, in \pp{} and \pA{} collisions the relative amount of particles from the \textit{bulk} (underlying event) is small, and therefore the characterisation of the jet fragmentation in small collision systems is necessary to accurately subtract non-flow effects and quantify potential biases in flow measurements. Several methods for removing non-flow contributions from measurements of $v_n$ rely on a subtraction of correlations in low-multiplicity events from those in high multiplicity collisions, and incorporate different assumptions about the multiplicity dependence of non-flow correlations. For example, a straight subtraction of the correlation functions measured in low-multiplicity events from those in high-multiplicity events is sufficient for extracting flow correlations, assuming that non-flow does not depend significantly on multiplicity and the low-multiplicity events are dominated by non-flow correlations~\cite{ALICE:2012eyl}. Meanwhile, template fit methods~\cite{ATLAS:2014qaj,ATLAS:2015hzw,Khachatryan:2016txc} allow for some multiplicity-dependence of non-flow by scaling the low-multiplicity correlation function before subtraction, but typically rely on alternative assumptions in order to constrain the scaling factors. Thus, an investigation into the multiplicity-dependence of the near-side jet peak, and therefore non-flow correlations, is important for non-flow subtraction techniques.

In the hot and dense environment of a heavy-ion collision, the products of hard (high momentum transfer) parton--parton scatterings, known as \textit{jets}, undergo strong interactions with the medium. The result is that the jet energy is redistributed to low-momentum particles, leading to a reduction in the number of high-\pt{} reconstructed jets and a modification of the momentum distribution of the jet fragments in \AAA{} collisions compared to \pp{} collisions, a phenomenon known as \textit{jet quenching}~\cite{Gyulassy:1999ig, Majumder:2010qh, Cunqueiro:2021wls}. The first demonstration of this jet modification with two-particle correlations was observed in \AuAu{} collisions measured by the STAR experiment at RHIC~\cite{STAR:2003pjh}. Further studies of hadron--hadron and jet--hadron correlations at RHIC~\cite{PHENIX:2005zfm,PHENIX:2006gto,PHENIX:2007yjc,STAR:2010mfd,STAR:2013thw,STAR:2016jdz} and the LHC~\cite{ALICE:2011gpa,CMS:2016qnj,CMS:2015hkr,ALICE:2016wkc,ALICE:2016mfw,ALICE:2016gso,ALICE:2018gni} have been used to quantitatively examine the jet structure in both \pp{} and \AAA{} collisions to gain insight into the mechanisms of jet quenching in heavy-ion collisions (for a review, see Ref.~\cite{Connors:2017ptx}). Models which do not include a hot and dense medium like the QGP are currently unable to fully describe the jet suppression and momentum redistribution of jet fragments observed in nucleus--nucleus collisions~\cite{Bierlich:2018xfw, daSilva:2020cyn}.  

Anisotropic flow, which describes the bulk evolution of the medium, and jet quenching, which represents the interactions between the hard and soft components of the collision, are typically viewed as signatures of the strongly-interacting medium produced in heavy-ion collisions~\cite{Gyulassy:1990ye,Wang:1991xy}. However, while flow-like signals have been measured in small collision systems, the effects of jet quenching are expected to be small in \pp{} and \pPb{} collisions~\cite{Zhang:2013oca, Tywoniuk:2014hta, Park:2016jap, Huss:2020whe, Huss:2020dwe, Zakharov:2021uza, Ke:2022gkq}, and have not been observed within experimental uncertainties~\cite{Adam:2014qja, Khachatryan:2016odn, Adam:2016jfp, Adam:2016dau, Acharya:2017okq,ALICE:2023plt,ATLAS:2022iyq,ALICE:2023ama,ALICE:2021wct,ALICE:2022qxg}.
Experimental and theoretical efforts are essential to understand these two observations within the same underlying physics picture in collisions of small nuclei.

As an alternative approach to jet suppression measurements, in this work the angular distributions and per-trigger associated yields from jet fragmentation are studied as a function of charged-particle multiplicity in pp collisions at the centre-of-mass energy of $\sqrt{s}=13$~TeV.
Thus, this work is a complementary search for jet quenching effects in small collision systems, in which the modification of the jet fragmentation is quantified as a function of multiplicity and compared to MC models. Model comparisons are necessary to draw conclusions on whether the multiplicity-dependence of the jet fragmentation, examined via the yield of jet particles and the pseudorapidity width, are due to kinematic selections or presence of the QGP medium.
Furthermore, this work provides input for studies of anisotropic flow by measuring the presence of non-flow as a function of relative pseudorapidity.

This article is organised into the following sections. First, the experimental setup and analysis method are described in Sec.~\ref{sec:experiment} and Sec.~\ref{sec:ana}, respectively. Section~\ref{sec:uncertainties} discusses the systematic uncertainties. The results and their comparison with model calculations are presented and discussed in Sec.~\ref{sec:results}. Finally, the results are summarised in~Sec.~\ref{sec:summary}.

%% file: sec_ana.tex

\section{Experimental setup and data samples}
\label{sec:experiment}

Here a brief description of the ALICE detector is given while a more comprehensive illustration of the ALICE detector and its performance can be found in Refs.~\cite{Aamodt:2008zz, Abelev:2014ffa, ALICE:2022wpn}. The analysis utilises the V0 detector~\cite{Abbas:2013taa}, the Inner Tracking System (ITS)~\cite{aliceITS}, and the Time Projection Chamber (TPC)~\cite{aliceTPC}.
The V0 detector consists of two arrays located on both sides of the interaction point, named V0A and V0C, each comprising 32 plastic scintillator tiles, having full azimuthal acceptance within the pseudorapidity intervals $2.8 < \eta < 5.1$ and $-3.7 < \eta < -1.7$, respectively. The ITS is a silicon tracker with six layers of silicon sensors. The two innermost layers of the ITS are called the Silicon Pixel Detector (SPD)~\cite{Santoro2009:ALICESPD}. The middle two layers are the Silicon Drift Detector, and the two outermost layers are the Silicon Strip Detector. The TPC is a gas-filled cylindrical tracking detector providing up to 159 reconstruction points for charged-particle tracks traversing the entire radial extent of the detector. The TPC and ITS have a pseudorapidity coverage of $-0.8<\eta<0.8$ and $-0.9<\eta<0.9$, respectively. 

The V0 provides a minimum bias (MB) trigger in pp collisions. A time coincidence of V0A and V0C signals triggers the data collection. The amplitudes of V0A and V0C signals are proportional to charged-particle multiplicity and their sum is denoted as V0M. The V0M amplitude is utilised to define the multiplicity class. In addition to V0M, the multiplicity can be classified using the number of clusters in the outer layer of the SPD, which covers the acceptance $|\eta|<$~1.4 to measure the number of produced charged tracks at midrapidity, $N_{\mathrm{ch}}$~\cite{ALICE:2020swj,ALICE:2023ulm, ALICE:2022kol}. The $N_{\mathrm{ch}}$ determined by the use of SPD clusters is corrected for detector acceptance as well as reconstruction and selection efficiency. The present paper utilises these two different multiplicity estimators covering different pseudorapidity ranges.

The analysed data samples of MB pp events at $\sqrt{s}=$~13 TeV were recorded during the LHC Run 2 period from 2016 to 2018 and correspond to an integrated luminosity ($\mathcal{L}_\mathrm{int}$) of about 19~nb$^{-1}$~\cite{ALICE-PUBLIC-2021-005}. The positions of primary vertices are reconstructed from track segments measured by the SPD. The reconstructed primary vertices are required to be within 8~cm of the nominal interaction point along the beam direction. Pileup events are identified as the events with multiple reconstructed primary vertices. These events are rejected if the longitudinal distance between any of the vertices to the primary vertex is greater than 0.8~cm. The relative rate of residual pileup events is estimated to range from $10^{-3}$ to $10^{-2}$ for MB events in pp collisions~\cite{ALICE:2020swj}. 

Charged-particle tracks are reconstructed in the pseudorapidity range $|\eta|<$~0.9 over the full azimuth with the TPC and ITS detectors, which are located inside a large solenoidal magnet, providing a uniform magnetic field of 0.5~T oriented along the beam axis. To guarantee good track momentum resolution, the reconstructed tracks are required to have crossed at least 70 readout pad rows in the TPC and have at least two associated hits in the ITS, with at least one in the SPD. The distances of closest approach (DCA) of thALICE:2016wkce track to the primary vertex in the longitudinal ($d_\mathrm{z}$) and transverse ($d_\mathrm{xy}$) directions are required to be $|d_\mathrm{z}|<$~2~cm and $|d_\mathrm{xy}|<$~(0.0105~$+$~0.0350~$\times p_{\mathrm{T}}^{-1.1})$~cm (with $p_{\mathrm{T}}$ in GeV/$c$), respectively, to suppress contamination from secondary charged particles originating from weakly decaying hadrons and interactions with the material. Due to the kinematics boosts that act in the longitudinal direction and geometry of SPD pixels, DCA cuts are kept looser in the longitudinal direction. The contamination is strongest at low \pt~and decreases from 9\% for $\pt<1$~GeV/$c$ to less than 1\% for $\pt>10$~GeV/$c$~\cite{ALICE:2018vuu}. The efficiency of charged-particle reconstruction is approximately 65\%~\cite{Ivanov:2006yra} at $\pt \sim$~0.15~GeV/$c$ and increases to about 80\% for particles with $\pt>1$ GeV/$c$. The $\pt$ resolution is approximately 1\% for primary charged particles~\cite{ALICE-PUBLIC-2017-005} with $\pt<$~1~GeV/$c$, and it linearly increases to 6\% at $\pt \sim$~50~GeV/$c$ in pp collisions~\cite{ALICE:2018vuu}. 
In this article, \textit{charged-particle} refers to a primary charged particle with a mean proper lifetime greater than 1~cm/$c$. This particle is either produced directly in the interaction or results from the decay of particles with a lifetime shorter than 1~cm/$c$, limited to decay chains leading to the initial collision (see Ref.~\cite{ALICE-PUBLIC-2017-005}).

\section{Analysis procedure}
\label{sec:ana}
The two-particle angular correlations are measured as a function of the relative azimuthal angle ($\Delta\varphi$) and the relative pseudorapidity ($\Delta\eta$) of two charged particles, referred to as trigger and associated particles, and is divided by the number of trigger particles $N_{\mathrm{trig}}$. The per-trigger yield is corrected for pair acceptance $B(\Delta\eta,\Delta\varphi)$, which is calculated by associating trigger particles in a given event with particles from other events (event-mixing). 
The acceptance-corrected two-particle angular correlation can be expressed as
\begin{equation}
\label{eq:twocorr}
\frac{1}{N_{\mathrm{trig}}}\frac{\mathrm{d}^{2}N_{\mathrm{pair}}}{\mathrm{d}\Delta\eta\mathrm{d}\Delta\varphi} = B(0,0)\frac{S(\Delta\eta,\Delta\varphi)}{B(\Delta\eta,\Delta\varphi)} \Big\lvert_{\pttrig,\,\ptassoc},
\end{equation}
where the number of trigger and associated particle pairs is denoted as $N_\mathrm{pair}$. The transverse momenta of trigger and associated particles are denoted as $\pttrig$ and $\ptassoc$, where $\pttrig$ is required to be higher than $\ptassoc$. All tracks are within $|\eta|<0.9$. The pair yield measured in the same event is represented by $S(\Delta\eta,\Delta\varphi)$. The longitudinal positions of the primary vertices of events to be mixed are required to be within the same, 2 cm wide, $z_{\rm{vtx}}$ interval for each multiplicity class used in this analysis.

The final correlation function is calculated by taking the weighted average of Eq.~\eqref{eq:twocorr} results obtained in these individual bins~\cite{Kopylov:1974th,ALICE:2016wkc}. In addition, all primary tracks are corrected for the single-particle tracking efficiency in a given multiplicity class as a function of $p_\mathrm{T}$, $\eta$, and $z_\mathrm{vtx}$. The corrections for tracking efficiency and acceptance are constructed with MC simulations and are based on events generated with PYTHIA 8.3 with the Monash tune~\cite{Skands:2014pea} and the GEANT3 transport package~\cite{Brun:1994aa}, which is employed to simulate the detector response.

\begin{figure}[!hbt]
\centering
    \includegraphics[width=\textwidth,trim={1.8cm 0 2.3cm 0},clip]{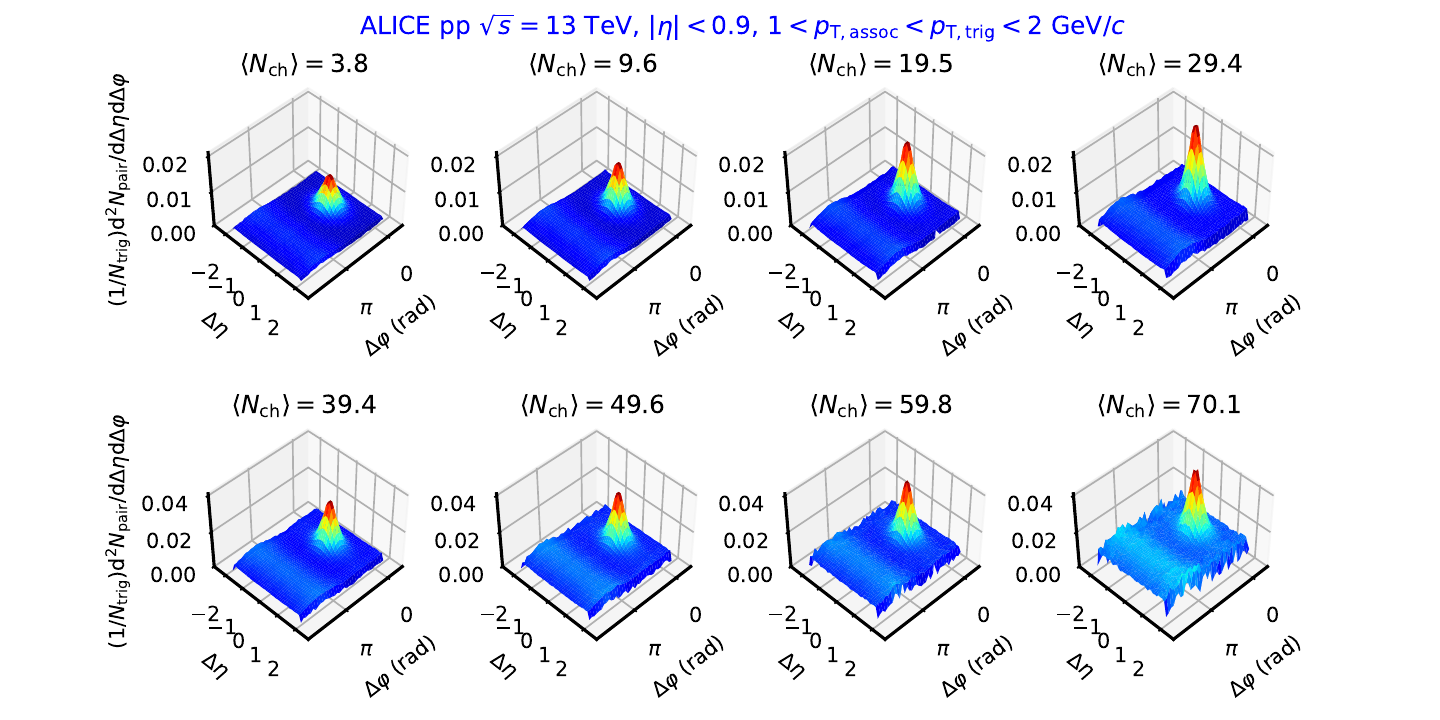} 
\caption{Two-particle angular correlation functions as a function of $\Delta\eta$ and $\Delta\varphi$ in pp collisions at $\sqrt{s}=13$ TeV for various multiplicity classes selected with the Midrapidity Multiplicity Estimator and characterised by the mean number of corrected tracks, $\langle N_{\mathrm{ch}}\rangle$ ($|\eta| < 1.0, \, \pt > 0.2~ \mathrm{GeV/}c$), with the minimum-bias multiplicity at $\langle N_\mathrm{ch}\rangle=11.3$. All correlation functions are shown for 1~$<\it{p}_{\rm{T,~assoc}}<\it{p}_{\rm{T,~trig}}<$~2~GeV/$c$.}
\label{fig:doubleridge}
\end{figure}

Figure~\ref{fig:doubleridge} presents the fully corrected two-particle angular correlation functions in pp collisions at $\sqrt{s}=13$ TeV for different multiplicity classes, which are classified with the mean number of reconstructed charged tracks at midrapidity ($\langle N_{\mathrm{trk}} \rangle$), referred to as the \textit{Midrapidity Multiplicity Estimator}. In order to correct for the effects of the tracking efficiency on the average multiplicity $\langle N_{\mathrm{ch}} \rangle$, a correlation between the reconstructed and simulated multiplicities in the simulated samples is formed, which is then randomly sampled to obtain a new multiplicity value from the generated distribution corresponding to the reconstructed value~\cite{ALICE:2023ulm, ALICE:2022kol}. The prominent near-side jet fragmentation peak around $(\Delta\eta,\Delta\varphi)\sim (0,0)$
is mostly dominated by particle pairs coming from the same jet while the away-side region ($\Delta\varphi\sim\pi$ and extended in the $\Delta\eta$ direction) is populated mostly by back-to-back jet correlations.

\begin{figure}[!hbt]
\centering
    \includegraphics[width=0.49\textwidth]{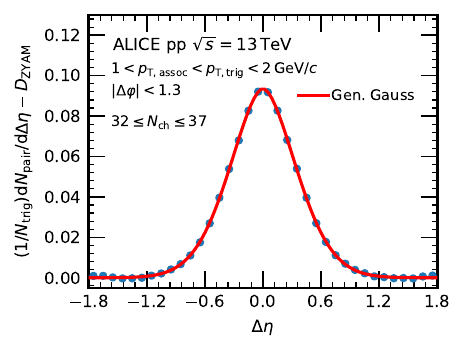} 
\caption{Near-side one-dimensional $\Delta\eta$ correlation function in pp collisions at $\sqrt{s}=$~13~TeV obtained using Eq.~\eqref{eq:etaproj}. The correlation is fitted with a generalised Gaussian distribution parameterised by Eq.~\eqref{eq:gaus}. The projected correlation function is presented after the baseline subtraction (see text for details). The statistical uncertainties are smaller than the marker size, and the systematic uncertainties are not shown.}
\label{fig:deltaetaproj}
\end{figure}

The shape of the jet fragmentation peak is quantified through a projection of the correlation function on to its $\Delta\eta$-axis
\begin{equation}
	\frac{1}{N_{\rm{trig}}} \frac{ \rm{d}\it{}N_{\rm{pair}} }{ \rm{d}\Delta\eta }=\int_{|\Delta\varphi|<1.3}\left( \frac{1}{\it{N}_{\rm{trig}}} \frac{ \rm{d}\it{}^{2} N_{\rm{pair}} }{ \rm{d}\Delta\eta d\Delta\varphi} \right)\dfrac{1}{\delta_{\Delta\varphi}} {\rm{d}}\Delta \varphi,
\label{eq:etaproj}
\end{equation}
where $\delta_{\Delta\varphi}=2.6$ is the normalisation constant for the projection range, which is also used in a previous publication~\cite{ALICE:2016wkc}. In order to characterise the near-side peak shape, the projected distribution is fitted with the generalised Gaussian function as shown in Fig.~\ref{fig:deltaetaproj}. This is pragmatic choice to characterise the shape without direct physics implications. The generalised Gaussian is given by
\begin{eqnarray}\label{eq:gaus}
    G_{\gamma_x,w_x}(x)=\frac{\gamma_x}{2{w}_x\Gamma (1/\gamma_x)} \exp \left[ -\left(\frac{|x|}{w_x}\right)^{\gamma_x} \right]+A,
\end{eqnarray}
where $w_x$ is the scale and $\gamma_x$ the shape parameter. These parameters can be used to quantify the width of the near-side peak and provide information on the jet fragmentation pattern. Furthermore, $\Gamma$ represents the Gamma function and $A$ is a constant background pedestal.
It is worth mentioning that $\gamma_x=$~2 and $\gamma_x=$~1 return Gaussian and Laplace distributions, respectively. The standard deviation ($\sigma$) of the generalised Gaussian function can be analytically expressed as $\sigma^2 \equiv w_x^{2} \Gamma(3/\gamma_x) / \Gamma(1/\gamma_x)$. Compared to the ordinary Gaussian, the generalised Gaussian function provides a better description of the projected correlation functions, especially in the region of large $\Delta\eta$ tails, which allows one to fit over the entire $\Delta\eta$-range.
Compared to the jet fragmentation peak, the small contribution of the near-side long-range correlations to the characterisation of its shape is considered negligible.

The near-side per-trigger associated yield ($Y^\mathrm{near}_{\rm{frag}}$) is measured by directly integrating the measured short-range $\Delta\eta$ correlations over $|\Delta\eta|<$~1.3~\cite{ALICE:2021nir,ALICE:2023ulm}, which can be expressed as
\begin{eqnarray}
Y^\mathrm{near}_{\rm{frag}} = \int_{|\Delta \eta|<1.3} \left( \frac{1}{\it{N}_{\rm{trig}}} \frac{ \rm{d}\it{}N_{\rm{pair}} }{ \rm{d}\Delta\eta }\right) \rm{d} \Delta\eta - \delta_{\Delta\eta}\,\it{D}_\mathrm{ZYAM}.
\label{eq:Ynear}
\end{eqnarray}
$D_\mathrm{ZYAM}$ defines the baseline of the ZYAM (Zero-Yield-at-Minimum) background subtraction~\cite{Ajitanand:2005jj}, and is taken to be equal to the pedestal term of the fit, i.e. $D_\mathrm{ZYAM}=A$, where $A$ is from Eq.~\eqref{eq:gaus}.

The same analysis procedure is used also for the other event-class definition used in this analysis based on the V0M amplitude, referred to as the \textit{Forward Multiplicity Estimator}. The mean charged-particle multiplicity density ($\langle N_{\mathrm{ch}} \rangle$) within $|\eta|<$~1.0 and $\pt>$~0.2~GeV/$c$ for a given V0M percentile class is derived based on the corresponding $\langle N_{\mathrm{ch}} \rangle$ for $|\eta|<$~0.5 and $\pt>$~0~GeV/$c$ from Ref.~\cite{ALICE:2020swj}. This value is then scaled to match the kinematic ranges of the particles at midrapidity using PYTHIA 8 with the Monash tune~\cite{Skands:2014pea} (the comparison can be found in Ref.~\cite{Ji:2023eqn}). This was repeated for all V0M amplitude percentiles. These values are corrected for acceptance and tracking efficiency, as well as for contamination by secondary particles.

\section{Systematic uncertainties}
\label{sec:uncertainties}

The systematic uncertainties are evaluated by varying different selection criteria including the primary vertex position range, DCA selection range, track selection criteria, constraints on the particle charges, and efficiency correction, and looking for the deviations of the extracted observables with respect to the default setup.
Unlike in the measurements of long-range correlations~\cite{ALICE:2021nir,ALICE:2023ulm}, the $\Delta\eta$ and $\Delta\varphi$ ranges are not varied, as these ranges are in the definition of the observable.

The sensitivity to detector acceptance effects is evaluated by varying the primary vertex position range for the event selection from $|z_\mathrm{vtx}|<8\,\mathrm{cm}$ to $|z_\mathrm{vtx}|<10\,\mathrm{cm}$. This effect is very small, of the order of 0--2\% for the majority of the observables. The uncertainty emerging from the secondary track contributions is estimated by applying a tighter DCA selection in the beam direction, the DCA limits were changed from
2 cm to 0.3 cm.
The resulting systematic uncertainty ranges from 0\% to 2\%. Different track selection criteria were tested.
The uncertainty related to this choice, referred to as tracking mode, is obtained by changing track selection criteria to those used in Refs.~\cite{ALICE:2019qyj,ALICE:2022jbp}, which yield a more uniform azimuthal distribution of tracks. 
This uncertainty is found to be mostly between 1\% to 3\%. The possible impact of the contamination by resonance decays on the near-side peak width is tested by correlating two particles with the same charge (like-sign pair). The corresponding relative deviations of the observables between the default and like-sign pair analysis is below 3\% at low $p_{\mathrm{T}}$ ($p_{\mathrm{T}}<$~1.5~GeV/$c$) and increases up to 5--10\% at high $p_{\mathrm{T}}$ ($p_{\mathrm{T}}\sim$~6~GeV/$c$). Finally, the systematic uncertainty from the efficiency correction is evaluated from simulated data by comparing the results obtained with MC generated-level particles with those obtained with the corrected reconstruction-level particles. The resulting uncertainty is broadly around 5\% for $p_{\mathrm{T}}<$~1.5~GeV/$c$ and increases up to 10\% at $p_{\mathrm{T}}\sim$~6~GeV/$c$ or multiplicity $\langle N_{\rm ch}\rangle\gtrsim 50$. The relative uncertainty resulting from the discrepancies in this closure test is the only source of uncertainty that differs significantly between the near-side jet width and yield observables, with the uncertainty on the width being typically much lower by 1--3\% in all bins. The total systematic uncertainty is calculated by summing up each source in quadrature assuming no correlations between them. The total systematic uncertainty ranges between 3\% and 15\%, with mostly smaller values being achieved for lower $p_\mathrm{T}$ bins.

%% file: sec_results.tex

\section {Results}
\label{sec:results}

\begin{figure}[!hbt]
    \centering
    \includegraphics[width=\textwidth]{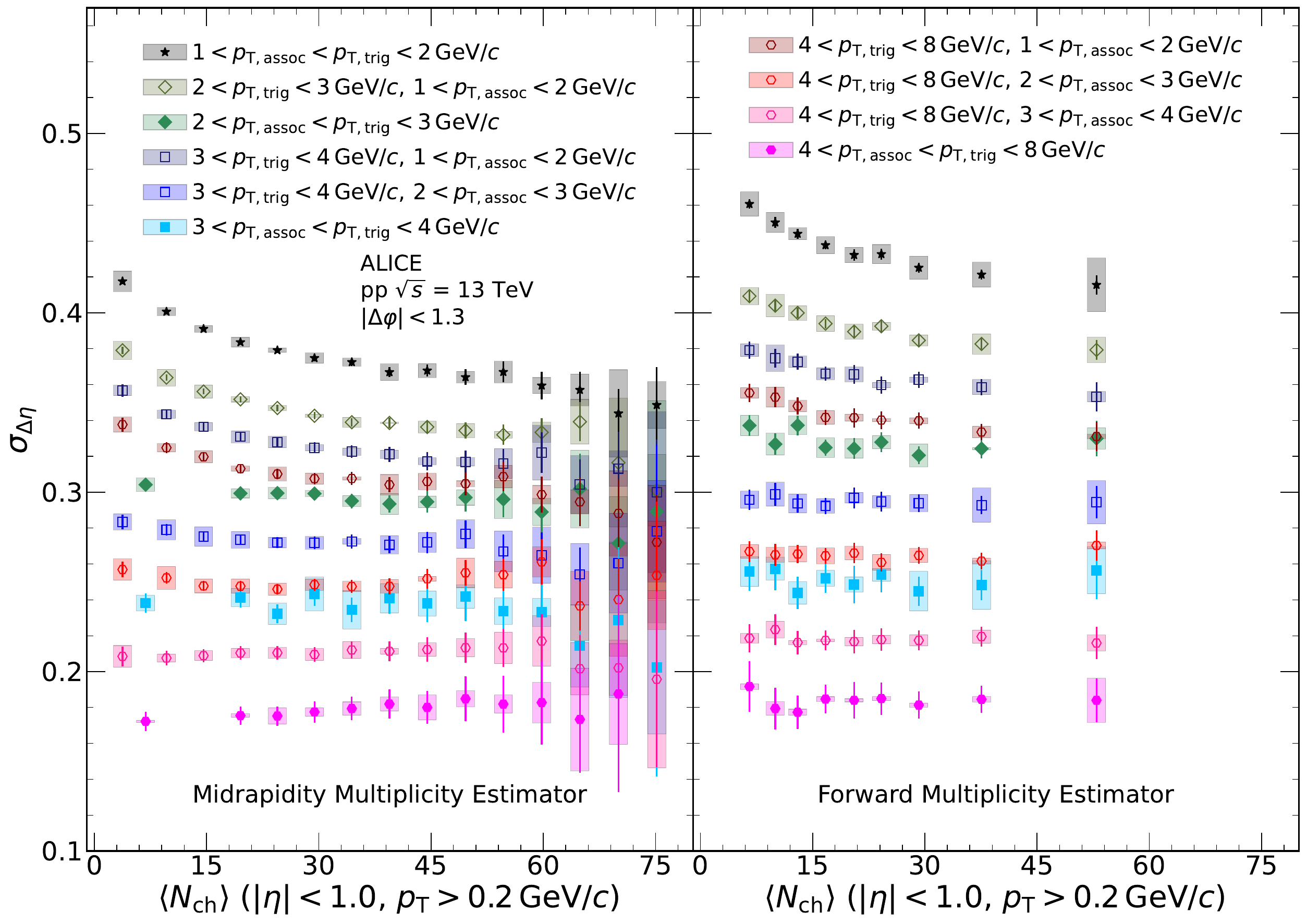}
    \caption{Near-side peak widths as a function of $\langle N_{\rm ch}\rangle$ for all \pTt~and~\pTa~intervals. Results with event activity measured at midrapidity (forward rapidity) are shown on the left (right). The statistical and systematic uncertainties are displayed with error bars and boxes, respectively.}
    \label{fig:jetshapemagnitude}
\end{figure}

Figure~\ref{fig:jetshapemagnitude} presents the width of the near-side peak as a function of charged-particle multiplicity in pp collisions at $\sqrt{s}=13$~TeV for various \pTt~and \pTa~intervals. The event activity is classified at midrapidity (left panel) and forward rapidity (right panel). For a given \pTt, the measured width decreases monotonously with increasing $\pTa$ due to the boost of jet fragments. The decreasing trend is observed for both multiplicity estimators using charged particles at midrapidity and forward rapidity. The peak width also decreases as a function of $\langle N_{\rm ch}\rangle$ for $\pTa<2$~GeV/$c$.

The narrowing trend of the near-side peak with increasing multiplicity is opposite to that observed in heavy-ion collisions for low $\pTt$ and $\pTa$ intervals, where a broadening of the near-side peak is measured~\cite{ALICE:2016wkc}. This happens only for the low associated-$\pt$ intervals ($\pt<2$~GeV/$c$), where the trend is strong.
This trend as a function of multiplicity cannot be attributed solely to jet-quenching effects, which in heavy-ion collisions are considered to produce the broadening of the near-side peak. There may be kinematic biases from the multiplicity selection, which could obscure the clear identification of a possible jet-quenching signature in pp collisions. Additionally, other comprehensive studies, which have used jet reconstruction methods~\cite{ALICE:2023plt, ALICE:2023oww}, have independently reached similar conclusions.

\begin{figure}[!hbt]
    \centering
    \includegraphics[width=\textwidth]{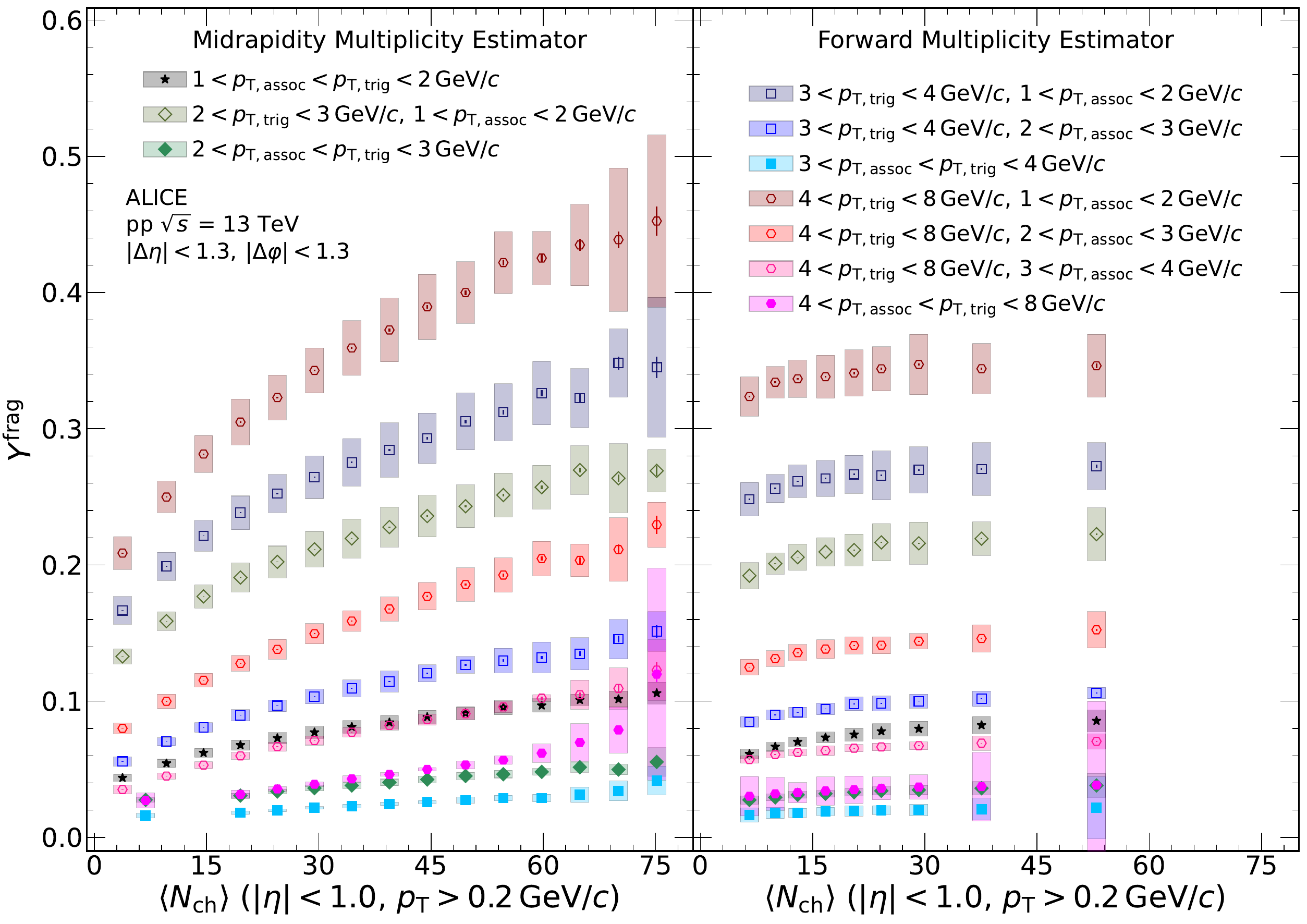}
    \caption{Per trigger normalised yield of associated particles in the near-side as a function of $\langle N_{\rm ch}\rangle$ for all $\pTt$ and $\pTa$ intervals. Results with event activity measured at midrapidity (forward rapidity) are shown on the left (right). The statistical and systematic uncertainties are displayed with error bars and boxes, respectively.}
    \label{fig:jetyieldmagnitude}
\end{figure}

Figure ~\ref{fig:jetyieldmagnitude} shows the near-side peak yields as a function of charged-particle multiplicity at midrapidity in pp collisions at $\sqrt{s}=13$~TeV for various \pTt~and \pTa~intervals and for the two event activity classifiers. 
The near-side yield increases with increasing multiplicity for both midrapidity and forward rapidity event multiplicity estimators. The trend increases more significantly as a function of the midrapidity than the forward rapidity multiplicity estimator due to auto-correlations~\cite{ALICE:2018pal}. 

\subsection{Comparisons with models}
\label{sec:theory}
\begin{figure}
    \centering
    \includegraphics[width=0.85\textwidth]{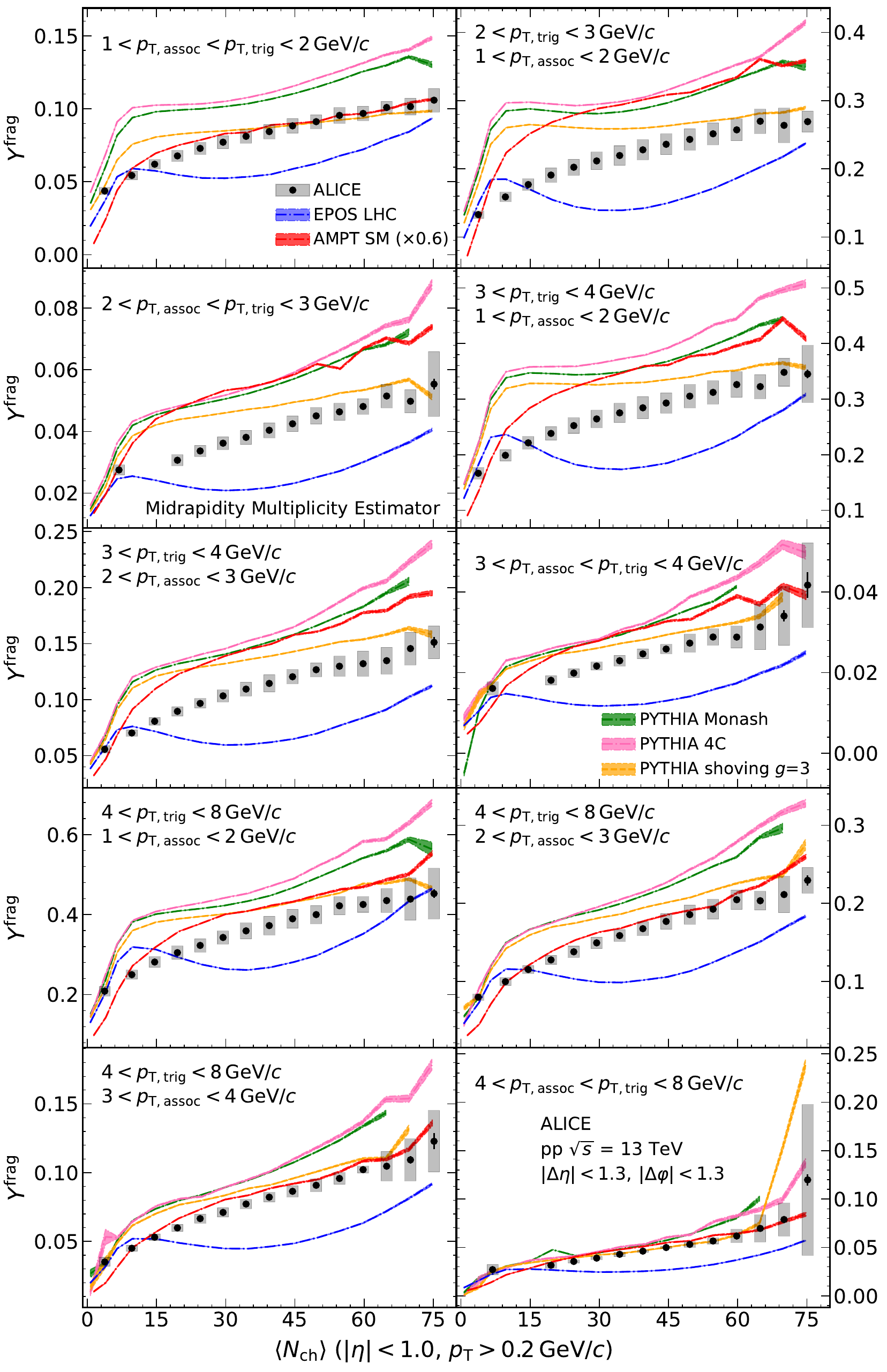}
    \caption{Multiplicity-dependence of near-side peak yields for all $p_\mathrm{T,\,trig}$ and $p_\mathrm{T,\,assoc}$ intervals measured in \pp{} collisions at $\sqrt{s} = 13$ TeV, compared with MC models. Multiplicities are estimated at midrapidity. The model results are shown as coloured bands with the width of the band denoting the statistical uncertainties. The error bars (boxes) represent the statistical (systematic) uncertainties in the experimental data. The AMPT calculations are scaled by a factor of 0.6 in all panels for visibility.}
    \label{fig:jetyieldFG}
\end{figure}

The measured data are compared with several MC event generators, namely PYTHIA 8, AMPT, and EPOS LHC.
PYTHIA 8~\cite{Bierlich:2022pfr} is a versatile event generator which can be used to simulate pp as well as heavy-ion collisions~\cite{Bierlich:2018xfw}.
Some parameters of the PYTHIA 8 model, mostly the ones associated with the non-perturbative regime of quantum chromodynamics, are tuned to reproduce experimental data. 
The default parameter set of PYTHIA 8 is called the Monash tune. It was adjusted based on a large set of LHC measurements and can describe the production of soft particles relatively well~\cite{Skands:2014pea}.
Its predecessor, the tune 4C, which is tuned to the first LHC measurements~\cite{Corke:2010yf}, is used in this analysis for comparison to the more recent versions.
In the default PYTHIA 8 versions, long-range correlations are not expected to emerge as there are no final-state partonic or hadronic interactions included. To describe these long-range correlations in HM pp collisions, string shoving can be enabled in PYTHIA 8~\cite{Bierlich:2017vhg,Bierlich:2019ixq}. This mechanism gives rise to a repulsive force acting between colour strings which yields a microscopic transverse pressure.
The string shoving approach in PYTHIA 8 reproduces the experimental measurements by ALICE~\cite{ALICE:2021nir} and CMS~\cite{Khachatryan:2016txc} of the long-range near-side ($|\Delta\varphi|<1.3$) ridge yield in high-multiplicity (HM) \pp\ events. 
One of the features of this model is that strings produced from hard scatterings are also affected by the repulsive force, which then leads to the observed long-range correlations even in low-multiplicity events with PYTHIA 8~\cite{Kim:2021elv}.

The EPOS model uses a core--corona approach to describe the evolution of heavy-ion collisions~\cite{Werner:2013tya}. The core is subject to hydrodynamic expansion, while the corona simulates the hadrons from string decays. Finally, UrQMD~\cite{Bass:1998ca, Bleicher:1999xi} is used to model hadronic interactions of all hadrons coming from the corona or the hadronisation of the core. 
A version called EPOS LHC, which includes a different parameterisation of flow in the case of a small and dense system, can successfully describe the long-range correlations in HM \pp{} events~\cite{ALICE:2021nir}. 

While models like EPOS~\cite{Werner:2013tya} use a causal hydrodynamic framework in describing the collective phenomena in small collision systems, the AMPT model with string melting~\cite{Lin:2004en} does this by modelling the evolution of the medium as a collection of interacting partons and hadrons~\cite{OrjuelaKoop:2015jss}.
The applicability of the model in reproducing flow results in small systems is studied in Ref.~\cite{Gallmeister:2018mcn}. 
The model can explain the long-range correlations for Pb--Pb collisions by introducing fluid-like and particle-like excitations with kinetic theory~\cite{Kurkela:2019kip,Kurkela:2020wwb,Ambrus:2021fej}. This study uses the same parton-interaction cross section value of 3 mb that is used in larger system studies~\cite{Lin:2004en}. This is important for the results to be comparable, as the partonic cross sections affect the final-state observables.

\begin{figure}
    \centering
    \includegraphics[width=0.85\textwidth]{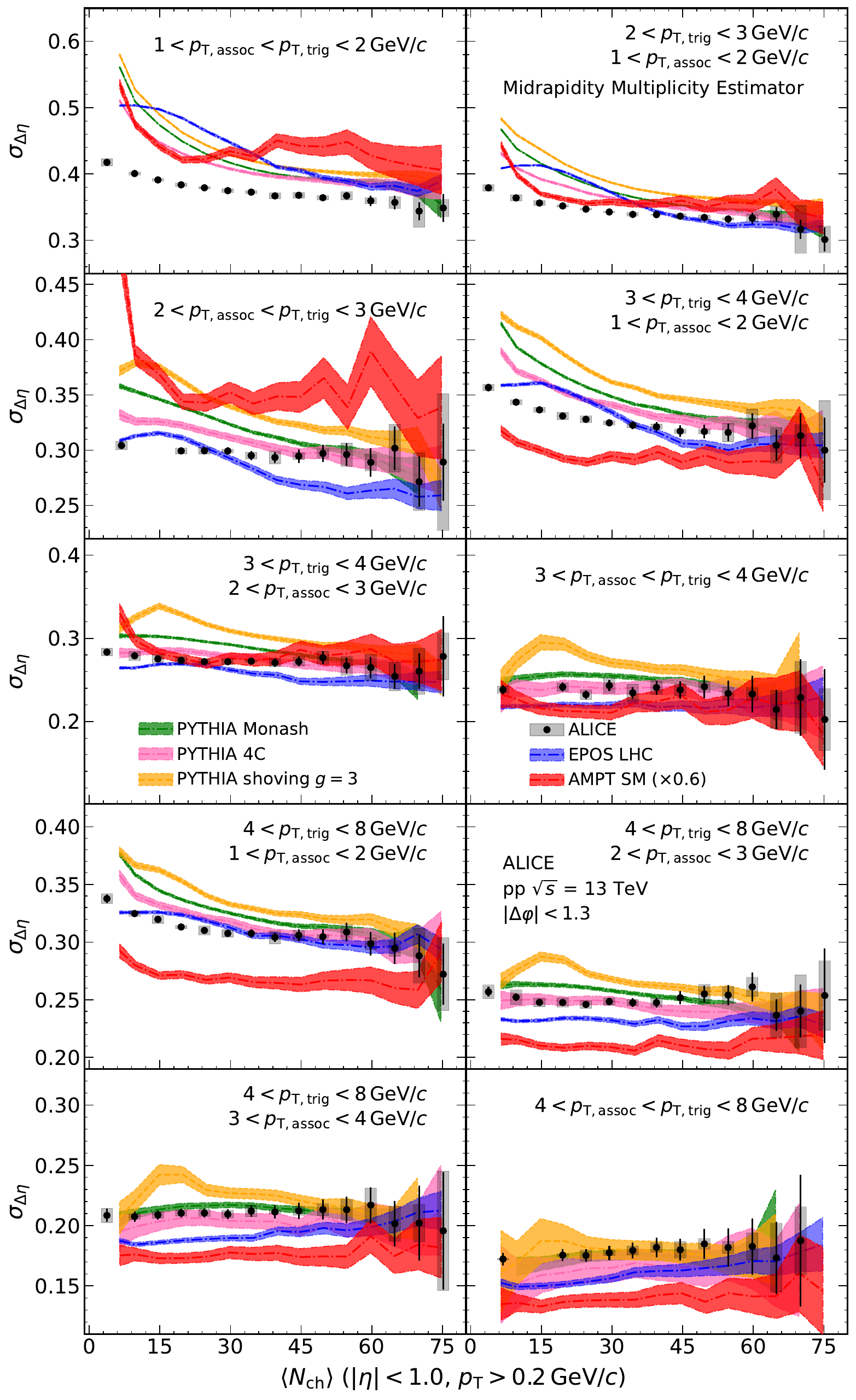}
    \caption{Multiplicity-dependence of near-side peak widths for all $p_\mathrm{T,\,trig}$ and $p_\mathrm{T,\,assoc}$ intervals measured in \pp{} collisions at $\sqrt{s} = 13$ TeV, compared with MC models. Multiplicities are estimated at midrapidity. The model results are shown as coloured bands with the width of the band denoting the statistical uncertainties. The error bars (boxes) represent the statistical (systematic) uncertainties in the experimental data. The AMPT calculations are scaled by a factor of 0.6 in all panels for visibility.}
    \label{fig:jetshapeFG}
\end{figure}

Figure~\ref{fig:jetyieldFG} displays MC model comparisons with the measured near-side peak yields for the multiplicity classes selected with the midrapidity estimator. The models reproduce qualitatively the growth of the yields as a function of multiplicity. However, in lower $p_\mathrm{T}$ intervals for $\langle N_{\rm ch} \rangle < 10$, the models predict a much steeper rise with increasing multiplicity. For the higher $p_\mathrm{T}$ intervals, the models agree better with the data. Similarly to the case of jet widths, the models overestimate the data for most $p_\mathrm{T}$ intervals with the exception of EPOS LHC, which underestimates the measured yields in all $\pt$ intervals. Only in the highest $p_\mathrm{T}$ interval, the different PYTHIA 8 tunes can describe the data within uncertainties. The AMPT model (scaled by a factor of 0.6 in all panels for visibility) overestimates the yields in all $\pt$ intervals , but captures the multiplicity dependence well in general.

Figure~\ref{fig:jetshapeFG} compares the measured near-side peak widths with the MC model expectations for the multiplicity selection based on the midrapidity estimator. In all $p_\mathrm{T}$ intervals, except the two with $\pTa > 3$ GeV/$c$ reported in the bottom panels of the figure, the simulated jet peaks get narrower with increasing multiplicity. The trend is stronger in the lower $p_\mathrm{T}$ intervals, while in the highest $p_\mathrm{T}$ intervals the multiplicity dependence becomes more flat. Interestingly, all models show similar multiplicity dependence with large overestimations of the near-side peak widths in the low multiplicity event classes. Furthermore, there is a subtle qualitative difference between the default PYTHIA 8 and PYTHIA 8 with string shoving, which suggests that the decreasing trend in the jet widths as a function of multiplicity might be connected to the final-state anisotropic particle distribution.
In addition, the multiplicity dependence becomes weaker for the higher-multiplicity events, which is more prominent at lower $\pt$. All the models describe qualitatively the trend of the data, but overestimate the measured near-side peak widths in lower multiplicity classes and low $p_\mathrm{T}$ intervals.
The largest discrepancy is observed for the AMPT model, which is for convenience scaled by a factor of 0.6 to fit in each of the panels. Yet, the AMPT model captures the multiplicity dependence of the near-side peak width. The PYTHIA 8 predictions and the data agree within uncertainties for $p_\mathrm{T,\, assoc}>3$~GeV/$c$, where the trend is flat.
Although, the model uncertainties are larger in the highest $\pt$ interval, the best description of the data is provided by the 4C-tune.
EPOS LHC presents a roughly similar magnitude and multiplicity dependence as the PYTHIA 8 models. 

For all the considered PYTHIA 8 tunes, the predictions agree better with the data at higher $p_\mathrm{T}$, specifically when {$p_\mathrm{T, \, assoc}>3$~GeV/$c$}. PYTHIA 8 tends to provide a better description of the data in higher $p_\mathrm{T}$ intervals because in this region the perturbation theory becomes more relevant. This shows that improvement is needed in the non-perturbative (lower-$p_\mathrm{T}$) regions.  These results can be used as critical constraints for further model improvements, especially in studying differences in the equilibrium and non-equilibrium descriptions~\cite{Kanakubo:2022ual,Romatschke:2017vte, Kurkela:2018wud, Kurkela:2018vqr, Kurkela:2018xxd, Strickland:2018ayk, Kamata:2020mka}. 

%% file: sec_summary.tex

\section{Conclusions}
\label{sec:summary}
The multiplicity dependence of short-range correlations for pairs of charged particles is measured with the technique of two-particle angular correlations to search for jet-quenching effects in pp collisions at ${\sqrt{{\textit s}}}=13$~TeV. The shape and yield extracted from short-range correlations are measured as a function of the relative azimuthal angle $\Delta\varphi$ and the pseudorapidity separation $\Delta\eta$ for pairs of primary charged particles within the pseudorapidity interval $|\eta| < 0.9$ and the transverse-momentum interval 1~$<p_{\mathrm{T,\,trig\,(assoc)}}<$~8~GeV/$c$.

The near-side peak width and per-trigger associated yield, which are both related to jet fragmentation patterns, are extracted by fitting and integrating the short-range $\Delta\eta$ correlations within $|\Delta\eta|<$~1.3 and $|\Delta\varphi|<$~1.3, in different multiplicity classes and $p_{\mathrm{T}}$ intervals. The near-side peak width monotonically decreases with increasing $p_{\mathrm{T}}$ due to the boost of evolving jet fragments. The near-side peak also gets narrower with increasing multiplicity for the low associated-$p_{\mathrm{T}}$ intervals ($p_{\mathrm{T}}<$~2 GeV/$c$). 
As discussed, the bias introduced by the measurement does not allow us to conclude on possible jet-fragmentation changes introduced by jet quenching.
The jet fragmentation yield is significantly dependent on the transverse momenta of trigger and associated particles and are found to increase with increasing multiplicity.

The experimental data are compared with calculations from three event generators. All generators describe qualitatively the decreasing trend of the near-side jet peak width as a function of charged particle density, which manifests itself in lower $p_{\mathrm{T}}$ intervals, suggesting that the decreasing trend is attributed to kinematic selections rather than collective-like effects.
Better descriptions are obtained from the models in higher $p_{\mathrm{T}}$ intervals, because there the system can be described more consistently with the perturbation theory. The narrowing trend of the near-side peak width toward HM events in experimental data and MC models suggests a potential bias in flow extraction methods that assume the independence of the near-side peak shape with multiplicity, such as the low-multiplicity (LM) template method~\cite{ALICE:2023lyr, ATLAS:2015hzw, ATLAS:2016yzd}. Except for the EPOS LHC shown at the lowest $p_\mathrm{T}$ interval, the near-side yield is overestimated by the other generators, although the considered PYTHIA 8 tunes describe the data better as $p_{\mathrm{T}}$ increases.

The measurement of the shape and yield of the near-side jet fragmentation peak is expected to provide a better understanding of particles emerging from jets and shows the narrowing trend of the peak width arising from kinematic selections in \pp{} collisions. Such findings will help constrain models, provide information on biases in flow extraction methods, and contribute to the search for jet-quenching effects in small collision systems.

%% file: fa_2024-08-25_Opt_C.tex

The ALICE Collaboration would like to thank all its engineers and technicians for their invaluable contributions to the construction of the experiment and the CERN accelerator teams for the outstanding performance of the LHC complex.
The ALICE Collaboration gratefully acknowledges the resources and support provided by all Grid centres and the Worldwide LHC Computing Grid (WLCG) collaboration.
The ALICE Collaboration acknowledges the following funding agencies for their support in building and running the ALICE detector:
A. I. Alikhanyan National Science Laboratory (Yerevan Physics Institute) Foundation (ANSL), State Committee of Science and World Federation of Scientists (WFS), Armenia;
Austrian Academy of Sciences, Austrian Science Fund (FWF): [M 2467-N36] and Nationalstiftung f\"{u}r Forschung, Technologie und Entwicklung, Austria;
Ministry of Communications and High Technologies, National Nuclear Research Center, Azerbaijan;
Conselho Nacional de Desenvolvimento Cient\'{\i}fico e Tecnol\'{o}gico (CNPq), Financiadora de Estudos e Projetos (Finep), Funda\c{c}\~{a}o de Amparo \`{a} Pesquisa do Estado de S\~{a}o Paulo (FAPESP) and Universidade Federal do Rio Grande do Sul (UFRGS), Brazil;
Bulgarian Ministry of Education and Science, within the National Roadmap for Research Infrastructures 2020-2027 (object CERN), Bulgaria;
Ministry of Education of China (MOEC) , Ministry of Science \& Technology of China (MSTC) and National Natural Science Foundation of China (NSFC), China;
Ministry of Science and Education and Croatian Science Foundation, Croatia;
Centro de Aplicaciones Tecnol\'{o}gicas y Desarrollo Nuclear (CEADEN), Cubaenerg\'{\i}a, Cuba;
Ministry of Education, Youth and Sports of the Czech Republic, Czech Republic;
The Danish Council for Independent Research | Natural Sciences, the VILLUM FONDEN and Danish National Research Foundation (DNRF), Denmark;
Helsinki Institute of Physics (HIP), Finland;
Commissariat \`{a} l'Energie Atomique (CEA) and Institut National de Physique Nucl\'{e}aire et de Physique des Particules (IN2P3) and Centre National de la Recherche Scientifique (CNRS), France;
Bundesministerium f\"{u}r Bildung und Forschung (BMBF) and GSI Helmholtzzentrum f\"{u}r Schwerionenforschung GmbH, Germany;
General Secretariat for Research and Technology, Ministry of Education, Research and Religions, Greece;
National Research, Development and Innovation Office, Hungary;
Department of Atomic Energy Government of India (DAE), Department of Science and Technology, Government of India (DST), University Grants Commission, Government of India (UGC) and Council of Scientific and Industrial Research (CSIR), India;
National Research and Innovation Agency - BRIN, Indonesia;
Istituto Nazionale di Fisica Nucleare (INFN), Italy;
Japanese Ministry of Education, Culture, Sports, Science and Technology (MEXT) and Japan Society for the Promotion of Science (JSPS) KAKENHI, Japan;
Consejo Nacional de Ciencia (CONACYT) y Tecnolog\'{i}a, through Fondo de Cooperaci\'{o}n Internacional en Ciencia y Tecnolog\'{i}a (FONCICYT) and Direcci\'{o}n General de Asuntos del Personal Academico (DGAPA), Mexico;
Nederlandse Organisatie voor Wetenschappelijk Onderzoek (NWO), Netherlands;
The Research Council of Norway, Norway;
Pontificia Universidad Cat\'{o}lica del Per\'{u}, Peru;
Ministry of Science and Higher Education, National Science Centre and WUT ID-UB, Poland;
Korea Institute of Science and Technology Information and National Research Foundation of Korea (NRF), Republic of Korea;
Ministry of Education and Scientific Research, Institute of Atomic Physics, Ministry of Research and Innovation and Institute of Atomic Physics and Universitatea Nationala de Stiinta si Tehnologie Politehnica Bucuresti, Romania;
Ministry of Education, Science, Research and Sport of the Slovak Republic, Slovakia;
National Research Foundation of South Africa, South Africa;
Swedish Research Council (VR) and Knut \& Alice Wallenberg Foundation (KAW), Sweden;
European Organization for Nuclear Research, Switzerland;
Suranaree University of Technology (SUT), National Science and Technology Development Agency (NSTDA) and National Science, Research and Innovation Fund (NSRF via PMU-B B05F650021), Thailand;
Turkish Energy, Nuclear and Mineral Research Agency (TENMAK), Turkey;
National Academy of  Sciences of Ukraine, Ukraine;
Science and Technology Facilities Council (STFC), United Kingdom;
National Science Foundation of the United States of America (NSF) and United States Department of Energy, Office of Nuclear Physics (DOE NP), United States of America.
In addition, individual groups or members have received support from:
Czech Science Foundation (grant no. 23-07499S), Czech Republic;
FORTE project, reg.\ no.\ CZ.02.01.01/00/22\_008/0004632, Czech Republic, co-funded by the European Union, Czech Republic;
European Research Council (grant no. 950692), European Union;
ICSC - Centro Nazionale di Ricerca in High Performance Computing, Big Data and Quantum Computing, European Union - NextGenerationEU;
Academy of Finland (Center of Excellence in Quark Matter) (grant nos. 346327, 346328), Finland.

%% file: Alice_Authorlist_2024-08-25_Opt_C.tex
\begin{flushleft} 
\small

S.~Acharya\,\orcidlink{0000-0002-9213-5329}\,$^{\rm 127}$, 
A.~Agarwal$^{\rm 135}$, 
G.~Aglieri Rinella\,\orcidlink{0000-0002-9611-3696}\,$^{\rm 32}$, 
L.~Aglietta\,\orcidlink{0009-0003-0763-6802}\,$^{\rm 24}$, 
M.~Agnello\,\orcidlink{0000-0002-0760-5075}\,$^{\rm 29}$, 
N.~Agrawal\,\orcidlink{0000-0003-0348-9836}\,$^{\rm 25}$, 
Z.~Ahammed\,\orcidlink{0000-0001-5241-7412}\,$^{\rm 135}$, 
S.~Ahmad\,\orcidlink{0000-0003-0497-5705}\,$^{\rm 15}$, 
S.U.~Ahn\,\orcidlink{0000-0001-8847-489X}\,$^{\rm 71}$, 
I.~Ahuja\,\orcidlink{0000-0002-4417-1392}\,$^{\rm 37}$
% ...

\section*{Affiliation Notes}

$^{\rm I}$ Deceased\\
$^{\rm II}$ Also at: Max-Planck-Institut fur Physik, Munich, Germany\\
$^{\rm III}$ Also at: Italian National Agency for New Technologies, Energy and Sustainable Economic Development (ENEA), Bologna, Italy\\
$^{\rm IV}$ Also at: Dipartimento DET del Politecnico di Torino, Turin, Italy\\
$^{\rm V}$ Also at: Yildiz Technical University, Istanbul, T\"{u}rkiye\\
$^{\rm VI}$ Also at: Department of Applied Physics, Aligarh Muslim University, Aligarh, India\\
$^{\rm VII}$ Also at: Institute of Theoretical Physics, University of Wroclaw, Poland\\
$^{\rm VIII}$ Also at: An institution covered by a cooperation agreement with CERN\\

\section*{Collaboration Institutes}

$^{1}$ A.I. Alikhanyan National Science Laboratory (Yerevan Physics Institute) Foundation, Yerevan, Armenia\\
$^{2}$ AGH University of Krakow, Cracow, Poland\\
$^{3}$ Bogolyubov Institute for Theoretical Physics, National Academy of Sciences of Ukraine, Kiev, Ukraine\\
$^{4}$ Bose Institute, Department of Physics  and Centre for Astroparticle Physics and Space Science (CAPSS), Kolkata, India\\
$^{5}$ California Polytechnic State University, San Luis Obispo, California, United States\\
$^{6}$ Central China Normal University, Wuhan, China\\
$^{7}$ Centro de Aplicaciones Tecnol\'{o}gicas y Desarrollo Nuclear (CEADEN), Havana, Cuba\\
$^{8}$ Centro de Investigaci\'{o}n y de Estudios Avanzados (CINVESTAV), Mexico City and M\'{e}rida, Mexico\\
$^{9}$ Chicago State University, Chicago, Illinois, United States\\
$^{10}$ China Institute of Atomic Energy, Beijing, China\\
$^{11}$ China University of Geosciences, Wuhan, China\\
$^{12}$ Chungbuk National University, Cheongju, Republic of Korea\\
$^{13}$ Comenius University Bratislava, Faculty of Mathematics, Physics and Informatics, Bratislava, Slovak Republic\\
$^{14}$ Creighton University, Omaha, Nebraska, United States\\
$^{15}$ Department of Physics, Aligarh Muslim University, Aligarh, India\\
$^{16}$ Department of Physics, Pusan National University, Pusan, Republic of Korea\\
$^{17}$ Department of Physics, Sejong University, Seoul, Republic of Korea\\
$^{18}$ Department of Physics, University of California, Berkeley, California, United States\\
$^{19}$ Department of Physics, University of Oslo, Oslo, Norway\\
$^{20}$ Department of Physics and Technology, University of Bergen, Bergen, Norway\\
$^{21}$ Dipartimento di Fisica, Universit\`{a} di Pavia, Pavia, Italy\\
$^{22}$ Dipartimento di Fisica dell'Universit\`{a} and Sezione INFN, Cagliari, Italy\\
$^{23}$ Dipartimento di Fisica dell'Universit\`{a} and Sezione INFN, Trieste, Italy\\
$^{24}$ Dipartimento di Fisica dell'Universit\`{a} and Sezione INFN, Turin, Italy\\
$^{25}$ Dipartimento di Fisica e Astronomia dell'Universit\`{a} and Sezione INFN, Bologna, Italy\\
$^{26}$ Dipartimento di Fisica e Astronomia dell'Universit\`{a} and Sezione INFN, Catania, Italy\\
$^{27}$ Dipartimento di Fisica e Astronomia dell'Universit\`{a} and Sezione INFN, Padova, Italy\\
$^{28}$ Dipartimento di Fisica `E.R.~Caianiello' dell'Universit\`{a} and Gruppo Collegato INFN, Salerno, Italy\\
$^{29}$ Dipartimento DISAT del Politecnico and Sezione INFN, Turin, Italy\\
$^{30}$ Dipartimento di Scienze MIFT, Universit\`{a} di Messina, Messina, Italy\\
$^{31}$ Dipartimento Interateneo di Fisica `M.~Merlin' and Sezione INFN, Bari, Italy\\
$^{32}$ European Organization for Nuclear Research (CERN), Geneva, Switzerland\\
$^{33}$ Faculty of Electrical Engineering, Mechanical Engineering and Naval Architecture, University of Split, Split, Croatia\\
$^{34}$ Faculty of Engineering and Science, Western Norway University of Applied Sciences, Bergen, Norway\\
$^{35}$ Faculty of Nuclear Sciences and Physical Engineering, Czech Technical University in Prague, Prague, Czech Republic\\
$^{36}$ Faculty of Physics, Sofia University, Sofia, Bulgaria\\
$^{37}$ Faculty of Science, P.J.~\v{S}af\'{a}rik University, Ko\v{s}ice, Slovak Republic\\
$^{38}$ Frankfurt Institute for Advanced Studies, Johann Wolfgang Goethe-Universit\"{a}t Frankfurt, Frankfurt, Germany\\
$^{39}$ Fudan University, Shanghai, China\\
$^{40}$ Gangneung-Wonju National University, Gangneung, Republic of Korea\\
$^{41}$ Gauhati University, Department of Physics, Guwahati, India\\
$^{42}$ Helmholtz-Institut f\"{u}r Strahlen- und Kernphysik, Rheinische Friedrich-Wilhelms-Universit\"{a}t Bonn, Bonn, Germany\\
$^{43}$ Helsinki Institute of Physics (HIP), Helsinki, Finland\\
$^{44}$ High Energy Physics Group,  Universidad Aut\'{o}noma de Puebla, Puebla, Mexico\\
$^{45}$ Horia Hulubei National Institute of Physics and Nuclear Engineering, Bucharest, Romania\\
$^{46}$ HUN-REN Wigner Research Centre for Physics, Budapest, Hungary\\
$^{47}$ Indian Institute of Technology Bombay (IIT), Mumbai, India\\
$^{48}$ Indian Institute of Technology Indore, Indore, India\\
$^{49}$ INFN, Laboratori Nazionali di Frascati, Frascati, Italy\\
$^{50}$ INFN, Sezione di Bari, Bari, Italy\\
$^{51}$ INFN, Sezione di Bologna, Bologna, Italy\\
$^{52}$ INFN, Sezione di Cagliari, Cagliari, Italy\\
$^{53}$ INFN, Sezione di Catania, Catania, Italy\\
$^{54}$ INFN, Sezione di Padova, Padova, Italy\\
$^{55}$ INFN, Sezione di Pavia, Pavia, Italy\\
$^{56}$ INFN, Sezione di Torino, Turin, Italy\\
$^{57}$ INFN, Sezione di Trieste, Trieste, Italy\\
$^{58}$ Inha University, Incheon, Republic of Korea\\
$^{59}$ Institute for Gravitational and Subatomic Physics (GRASP), Utrecht University/Nikhef, Utrecht, Netherlands\\
$^{60}$ Institute of Experimental Physics, Slovak Academy of Sciences, Ko\v{s}ice, Slovak Republic\\
$^{61}$ Institute of Physics, Homi Bhabha National Institute, Bhubaneswar, India\\
$^{62}$ Institute of Physics of the Czech Academy of Sciences, Prague, Czech Republic\\
$^{63}$ Institute of Space Science (ISS), Bucharest, Romania\\
$^{64}$ Institut f\"{u}r Kernphysik, Johann Wolfgang Goethe-Universit\"{a}t Frankfurt, Frankfurt, Germany\\
$^{65}$ Instituto de Ciencias Nucleares, Universidad Nacional Aut\'{o}noma de M\'{e}xico, Mexico City, Mexico\\
$^{66}$ Instituto de F\'{i}sica, Universidade Federal do Rio Grande do Sul (UFRGS), Porto Alegre, Brazil\\
$^{67}$ Instituto de F\'{\i}sica, Universidad Nacional Aut\'{o}noma de M\'{e}xico, Mexico City, Mexico\\
$^{68}$ iThemba LABS, National Research Foundation, Somerset West, South Africa\\
$^{69}$ Jeonbuk National University, Jeonju, Republic of Korea\\
$^{70}$ Johann-Wolfgang-Goethe Universit\"{a}t Frankfurt Institut f\"{u}r Informatik, Fachbereich Informatik und Mathematik, Frankfurt, Germany\\
$^{71}$ Korea Institute of Science and Technology Information, Daejeon, Republic of Korea\\
$^{72}$ KTO Karatay University, Konya, Turkey\\
$^{73}$ Laboratoire de Physique Subatomique et de Cosmologie, Universit\'{e} Grenoble-Alpes, CNRS-IN2P3, Grenoble, France\\
$^{74}$ Lawrence Berkeley National Laboratory, Berkeley, California, United States\\
$^{75}$ Lund University Department of Physics, Division of Particle Physics, Lund, Sweden\\
$^{76}$ Nagasaki Institute of Applied Science, Nagasaki, Japan\\
$^{77}$ Nara Women{'}s University (NWU), Nara, Japan\\
$^{78}$ National and Kapodistrian University of Athens, School of Science, Department of Physics , Athens, Greece\\
$^{79}$ National Centre for Nuclear Research, Warsaw, Poland\\
$^{80}$ National Institute of Science Education and Research, Homi Bhabha National Institute, Jatni, India\\
$^{81}$ National Nuclear Research Center, Baku, Azerbaijan\\
$^{82}$ National Research and Innovation Agency - BRIN, Jakarta, Indonesia\\
$^{83}$ Niels Bohr Institute, University of Copenhagen, Copenhagen, Denmark\\
$^{84}$ Nikhef, National institute for subatomic physics, Amsterdam, Netherlands\\
$^{85}$ Nuclear Physics Group, STFC Daresbury Laboratory, Daresbury, United Kingdom\\
$^{86}$ Nuclear Physics Institute of the Czech Academy of Sciences, Husinec-\v{R}e\v{z}, Czech Republic\\
$^{87}$ Oak Ridge National Laboratory, Oak Ridge, Tennessee, United States\\
$^{88}$ Ohio State University, Columbus, Ohio, United States\\
$^{89}$ Physics department, Faculty of science, University of Zagreb, Zagreb, Croatia\\
$^{90}$ Physics Department, Panjab University, Chandigarh, India\\
$^{91}$ Physics Department, University of Jammu, Jammu, India\\
$^{92}$ Physics Program and International Institute for Sustainability with Knotted Chiral Meta Matter (SKCM2), Hiroshima University, Hiroshima, Japan\\
$^{93}$ Physikalisches Institut, Eberhard-Karls-Universit\"{a}t T\"{u}bingen, T\"{u}bingen, Germany\\
$^{94}$ Physikalisches Institut, Ruprecht-Karls-Universit\"{a}t Heidelberg, Heidelberg, Germany\\
$^{95}$ Physik Department, Technische Universit\"{a}t M\"{u}nchen, Munich, Germany\\
$^{96}$ Politecnico di Bari and Sezione INFN, Bari, Italy\\
$^{97}$ Research Division and ExtreMe Matter Institute EMMI, GSI Helmholtzzentrum f\"ur Schwerionenforschung GmbH, Darmstadt, Germany\\
$^{98}$ Saga University, Saga, Japan\\
$^{99}$ Saha Institute of Nuclear Physics, Homi Bhabha National Institute, Kolkata, India\\
$^{100}$ School of Physics and Astronomy, University of Birmingham, Birmingham, United Kingdom\\
$^{101}$ Secci\'{o}n F\'{\i}sica, Departamento de Ciencias, Pontificia Universidad Cat\'{o}lica del Per\'{u}, Lima, Peru\\
$^{102}$ Stefan Meyer Institut f\"{u}r Subatomare Physik (SMI), Vienna, Austria\\
$^{103}$ SUBATECH, IMT Atlantique, Nantes Universit\'{e}, CNRS-IN2P3, Nantes, France\\
$^{104}$ Sungkyunkwan University, Suwon City, Republic of Korea\\
$^{105}$ Suranaree University of Technology, Nakhon Ratchasima, Thailand\\
$^{106}$ Technical University of Ko\v{s}ice, Ko\v{s}ice, Slovak Republic\\
$^{107}$ The Henryk Niewodniczanski Institute of Nuclear Physics, Polish Academy of Sciences, Cracow, Poland\\
$^{108}$ The University of Texas at Austin, Austin, Texas, United States\\
$^{109}$ Universidad Aut\'{o}noma de Sinaloa, Culiac\'{a}n, Mexico\\
$^{110}$ Universidade de S\~{a}o Paulo (USP), S\~{a}o Paulo, Brazil\\
$^{111}$ Universidade Estadual de Campinas (UNICAMP), Campinas, Brazil\\
$^{112}$ Universidade Federal do ABC, Santo Andre, Brazil\\
$^{113}$ Universitatea Nationala de Stiinta si Tehnologie Politehnica Bucuresti, Bucharest, Romania\\
$^{114}$ University of Cape Town, Cape Town, South Africa\\
$^{115}$ University of Derby, Derby, United Kingdom\\
$^{116}$ University of Houston, Houston, Texas, United States\\
$^{117}$ University of Jyv\"{a}skyl\"{a}, Jyv\"{a}skyl\"{a}, Finland\\
$^{118}$ University of Kansas, Lawrence, Kansas, United States\\
$^{119}$ University of Liverpool, Liverpool, United Kingdom\\
$^{120}$ University of Science and Technology of China, Hefei, China\\
$^{121}$ University of South-Eastern Norway, Kongsberg, Norway\\
$^{122}$ University of Tennessee, Knoxville, Tennessee, United States\\
$^{123}$ University of the Witwatersrand, Johannesburg, South Africa\\
$^{124}$ University of Tokyo, Tokyo, Japan\\
$^{125}$ University of Tsukuba, Tsukuba, Japan\\
$^{126}$ Universit\"{a}t M\"{u}nster, Institut f\"{u}r Kernphysik, M\"{u}nster, Germany\\
$^{127}$ Universit\'{e} Clermont Auvergne, CNRS/IN2P3, LPC, Clermont-Ferrand, France\\
$^{128}$ Universit\'{e} de Lyon, CNRS/IN2P3, Institut de Physique des 2 Infinis de Lyon, Lyon, France\\
$^{129}$ Universit\'{e} de Strasbourg, CNRS, IPHC UMR 7178, F-67000 Strasbourg, France, Strasbourg, France\\
$^{130}$ Universit\'{e} Paris-Saclay, Centre d'Etudes de Saclay (CEA), IRFU, D\'{e}partment de Physique Nucl\'{e}aire (DPhN), Saclay, France\\
$^{131}$ Universit\'{e}  Paris-Saclay, CNRS/IN2P3, IJCLab, Orsay, France\\
$^{132}$ Universit\`{a} degli Studi di Foggia, Foggia, Italy\\
$^{133}$ Universit\`{a} del Piemonte Orientale, Vercelli, Italy\\
$^{134}$ Universit\`{a} di Brescia, Brescia, Italy\\
$^{135}$ Variable Energy Cyclotron Centre, Homi Bhabha National Institute, Kolkata, India\\
$^{136}$ Warsaw University of Technology, Warsaw, Poland\\
$^{137}$ Wayne State University, Detroit, Michigan, United States\\
$^{138}$ Yale University, New Haven, Connecticut, United States\\
$^{139}$ Yonsei University, Seoul, Republic of Korea\\
$^{140}$ Affiliated with an institute covered by a cooperation agreement with CERN\\
$^{141}$ Affiliated with an international laboratory covered by a cooperation agreement with CERN.\\

\end{flushleft} 